\documentclass[11pt]{article}
\usepackage{amsfonts,amssymb,amsmath,amscd,cite,palatino}
\usepackage{geometry}                
\usepackage{graphicx}
\usepackage{amssymb}
\usepackage{epstopdf}
\def\be{\begin{equation}}
\def\beq{\begin{equation}}
\def\ee{\end{equation}}
\def\eeq{\end{equation}}
\def\ba{\begin{eqnarray}}
\def\ea{\end{eqnarray}}
\def\eqn#1{\begin{equation}\begin{split}#1\end{split}\end{equation}}

\def\pmat#1{\begin{pmatrix}#1\end{pmatrix}}

\allowdisplaybreaks[1]
\newtheorem{Ex}{Exercise}		
\newtheorem{Thm}{Theorem}	

\newcommand{\DD}{\mathcal{D}}
\newcommand{\EE}{\mathcal{E}}
\newcommand{\HH}{\mathcal{H}}	
\newcommand{\RR}{\mathcal{R}}	
\newcommand{\LD}{\mathcal{L}}				
\newcommand{\GLD}{\widehat{\LD}}		

\def\HH{{\cal H}}

\title{Double Field Theory, T-Duality, and Courant Brackets}
\author{Barton Zwiebach, MIT}
\date{Munich, August 2010}                                           

\begin{document}
\maketitle

\begin{abstract}
These lecture notes are based on three lectures, each ninety minutes long, given by the author during the ``International School on Strings and Fundamental Physics'' which took place in Garching/Munich from July 25 to August 6, 2010. These lectures, aimed at graduate students, require only a basic knowledge of string theory and give a simple introduction to
double field theory.  These notes were prepared by Marco Baumgartl and
Nicolas Moeller.
\end{abstract}

\hfill{MIT-CTP-4293}

\section{Introduction}

These lectures focus on making T-duality explicit in {\em field} theory Lagrangians.   The `T' in T-duality stands for `toroidal'. T-duality is an old and still fascinating topic in string theory. We will develop some Lagrangians for T-dual field theories that are quite intriguing and may have interesting applications.  The material covered here is based on joint work with Chris Hull and Olaf Hohm.  These notes are informal and do not attempt to be comprehensive nor to provide complete references. They deal with the basics of the subject and do not describe nor cite any of the recent developments.

Theories implementing T-duality bring up mathematical constructions 
such as the Courant-brackets as well as elements of generalized geometry. There is plenty of mathematical work on these topics, much of it in the context of first-quantized string theory. In our double 
field theory context, Courant-brackets and ideas of generalized geometry  appear in a very natural way and help construct the Lagrangians.  

Courant-brackets are natural generalizations of the Lie brackets that govern general relativity. Courant-brackets should be relevant to the effective field theory of strings and we are beginning to see this. Before entering this fascinating topic we will first talk about strings in toroidal backgrounds and some of their important properties.

\section{String theory in toroidal backgrounds}

Consider a closed string living in a spacetime with a compactified coordinates. It is well known that upon quantization there will be momentum modes and winding modes for each compact direction. Let us denote the compact coordinates by $x^a$ and the non-compact coordinates by $x^\mu$, with $x^i = (x^a, x^\mu)$. The compact coordinates $x^a$ give rise to string momentum excitations $p_a$. Since strings are extended objects, there are also  winding quantum numbers $w^a$. These should in fact be associated to some new coordinates $\tilde x_a$. If one attempts to write down the complete field theory of closed strings in coordinate space it will include the $x^a$ as well as the $\tilde x_a$. Thus, the arguments of all fields in such a theory will be doubled and we call it a double field theory (DFT).
The doubled fields $\phi(x^a, \tilde x_a, x^\mu)$ are said to be functions of momentum {\it and} winding.

Since the field arguments are doubled, actions must include a suitable integration over the additional dual coordinates:  
\eqn{
	S = \int dx^ad\tilde x_ad x^\mu{\cal L}(x^a, \tilde x_a, x^\mu)\ .
}

It is clear from the basic ideas of closed string field theory  that the full string theory is described by a Lagrangian of this form. With an infinite number of fields included, however, it is very complicated. A simplification can be achieved by restricting to a subset of fields only. The natural restriction is to consider only the massless sector, which includes a dilaton $\phi$, a metric $g_{ij}$ with Riemann curvature $R(g)$, and a Kalb-Ramond field $b_{ij}$ with field strength $H=db$.

The familiar low energy effective field theory of the bosonic closed string for these massless fields is given by
\eqn{\label{lowenergyaction}
	S_* &= \int dx\sqrt{-g}e^{-2\phi}\Bigl[ R + 4(\partial\phi)^2 - \frac{1}{12}H^2\Bigr]+ \dots
}
where the dots denote higher-derivative terms. In the light of the coordinate doubling on tori, what will this action become~?

There will be quite some obstacles in finding the correct action. One leading principle which helps in its construction is generalized geometry. Generalized geometry is in fact a very mild generalization of geometry. Let us look at its gauge symmetry first.
Its gauge symmetry parameters are vector fields $\xi^i \in T(M)$, which parametrize diffeomorphisms and live in the tangent bundle of the manifold, together with one-forms $\tilde\xi_i\in T^*(M)$, which describe gauge transformations of $b_{ij}$ and live in the dual tangent bundle. Both are combined naturally in the setup of generalized geometry, 
\eqn{
	\xi+\tilde\xi\in T(M)\oplus T^*(M)\ .
}
Generalized geometry does not double any coordinates. What it does achieve is to treat vectors and one-forms on an equal footing, so that it makes sense to add them to an object living in the sum of the tangent space and its dual.

In generalized geometry the Courant-bracket is the right extension of the Lie bracket. We will see that it will play a prominent role in our construction.
Also, in generalized geometry and string theory the 
field ${\cal E}_{ij}=g_{ij}+b_{ij}$ appears repeatedly, and 
one also has  
the generalized metric ${\cal H}^{MN}$. The generalized metric is a key structure also in string theory.  Up to now there 
were no actions written explicitly in terms of these fields.

In the following we will write down double field theories that are T-duality covariant versions of $S_*$. We will find that Courant-brackets, the field ${\cal E}_{ik}$, and the generalized metric ${\cal H}^{MN}$ will play an important role. 

\subsection{Sigma-model action}

In order to construct a first-quantized action, we start with the usual sigma-model action for strings propagating in a background. It is given by
\eqn{
\label{smaction}
	S &= -\frac{1}{4\pi}\int_0^{2\pi}d\sigma \int_{-\infty}^\infty d\tau \Bigl(
		\eta^{\alpha\beta}\partial_\alpha X^i\partial_\beta X^j G_{ij} + 
		\epsilon^{\alpha\beta}\partial_\alpha X^i\partial_\beta X^j B_{ij}
		\Bigr) \,,
}
where
\eqn{
	\eta^{\alpha\beta} &= \text{diag} (-1, 1), \qquad \epsilon^{01}=-1,
	~~~~\quad ~\partial_\alpha =  (\partial_\tau \,, \partial_\sigma)\,, 
	\\[0.3ex]
	X^i &= (X^a, X^\mu)\qquad X^a \sim X^a+2\pi, 
	\qquad i=0, \dots , D-1\,.
}
The $X^a$ are  periodic coordinates for the compact dimensions. The total number of dimensions is $D$.
The closed string background fields $G$ and $B$ are $D\times D$ matrices and are taken to be constant with the following properties:
\eqn{
	G_{ij} &= \begin{pmatrix}\hat G_{ab} & 0 \\ 0 & \eta_{\mu\nu}\end{pmatrix}\,, \qquad
	B_{ij} = \begin{pmatrix}\hat B_{ab} & 0 \\ 0 & 0 \end{pmatrix}\,, \qquad
	G^{ij}G_{jk} = \delta^i_k\,.
}
Both $G$ and $B$ can be combined into the field $E$ defined by
\eqn{
	E_{ij} &= G_{ij}+B_{ij} = \begin{pmatrix}\hat E_{ab} & 0 \\ 0 & \eta_{\mu\nu}\end{pmatrix}\,,
	\quad\text{with}\quad \hat E_{ab} = \hat G_{ab} + \hat B_{ab}\ .
}

\begin{Ex}
By using the action (\ref{smaction}), prove that the canonical momentum $P_i$ is given by
\eqn{
	2\pi P_i &= G_{ij}\dot X^j + B_{ij}{X'}^j\,,
}
(dot for $\partial_\tau$, prime for $\partial_\sigma$) and that the Hamiltonian density $\underline H$ takes the form 
\eqn{
\label{Hamiltonian}
	4\pi \underline H &= \begin{pmatrix}X'\,,& 2\pi P\end{pmatrix}  \; {\cal H}(E) \; \begin{pmatrix}X'\\2\pi P\end{pmatrix} \,,
}
with the $2D\times 2D$ matrix
\eqn{
\label{Hamiltonian2}
	{\cal H}(E) &= \begin{pmatrix} G- BG^{-1}B & BG^{-1}\\ -G^{-1}B & G^{-1}\end{pmatrix} \ .
}
\end{Ex}

The matrix ${\cal H}(E)$ is a $2D\times 2D$ symmetric matrix  constructed out of $G$ and $B$. It is called the `generalized metric.'
More precisely we will identify it with an object ${\cal H}^{MN}$ with $M, N=1, \dots, 2D$.
It is convenient to write ${\cal H}$ and its inverse in product form as
\eqn{
	{\cal H} &= \begin{pmatrix} G- BG^{-1}B & BG^{-1}\\ -G^{-1}B & G^{-1}\end{pmatrix}
		=
		\begin{pmatrix} 1 & B \\ 0 & 1 \end{pmatrix}
		\begin{pmatrix} G & 0 \\ 0 & G^{-1} \end{pmatrix}
		\begin{pmatrix} \phantom{-}1 & 0 \\ -B & 1 \end{pmatrix} \,,\\[0.5ex]
	{\cal H}^{-1} &= 
		\begin{pmatrix} G^{-1} & - G^{-1}B \\ BG^{-1} & G-BG^{-1}B\end{pmatrix}
		=
		\begin{pmatrix} 1 & 0 \\ B & 1 \end{pmatrix}
		\begin{pmatrix} G^{-1} & 0 \\ 0 & G \end{pmatrix}
		\begin{pmatrix} 1 & -B \\ 0 & \phantom{-}1 \end{pmatrix}\,.
}
${\cal H}$ is non-degenerate because each of its factors is non-degenerate. It is useful to define another metric $\eta$ with constant off-diagonal entries
\eqn{
\label{etaHeta}
	\eta = \begin{pmatrix}0 & 1\\1&0\end{pmatrix}\ . 
}
With the metric $\eta$ we are able to relate ${\cal H}$ and its inverse, so that, as you can check, 
\eqn{
\label{eq-0}
	\eta {\cal H} \eta & = {\cal H}^{-1} \ .
}
Such a constraint comes about because the generalized metric 
is a $2D\times 2D$ matrix symmetric matrix  constructed from a 
single $D\times D$ matrix $E= G+B$.  Thus is has to be constrained.
We can view the parameterization of ${\cal H}$ in terms of $G$ and $B$
as a natural and general solution of the constraint.

Let us put indices on ${\cal H}$ like on a metric, so that we can identify
\eqn{\label{oddnotation}
	{\cal H} & \ \leftrightarrow \ {\cal H}^{MN} \,,\\[0.3ex]
	{\cal H}^{-1} & \ \leftrightarrow \ {\cal H}_{MN}\,. \\
}
Then equation (\ref{eq-0}) becomes
\eqn{
	\eta_{PM}{\cal H}^{MN}\eta_{NQ} &= {\cal H}_{PQ} \\[0.3ex]
	{\cal H}^{MN}\eta_{MP}\eta_{NQ} &= {\cal H}_{PQ}\ ,
}
so that lowering the indices of $\HH$ with the metric $\eta$ gives us the inverse $\HH^{-1}$!  The capitalized indices $M,N$ run over $2D$ values,
and will be called $O(D,D)$ indices.

\subsection{Oscillator expansions}
The string coordinate $X^i = x^i + w^i\sigma + \tau G^{ij}p_j + \dots$ has an expansion in terms of momenta, winding, and oscillators. The zero modes $\alpha_0$ and $\tilde \alpha_0$ are given by
\eqn{
	\alpha_0^i &= \frac{1}{\sqrt{2}}G^{ij}(p_j - E_{jk}w^k) \,,\\
	\tilde \alpha_0^i &= \frac{1}{\sqrt{2}}G^{ij}(p_j + E_{kj}w^k)\,.
}
Written with $p_i ={1\over i} {\partial\over \partial x^i}$
and $w^i ={1\over i} {\partial\over \partial \tilde x_i}$
\eqn{
	\alpha_{0i} &= -\frac{i}{\sqrt{2}}\Bigl(\frac{\partial}{\partial x^i} - E_{ik}\frac{\partial}{\partial \tilde x_k} \Bigr) \equiv -\frac{i}{\sqrt{2}} D_i\,, \\
	\tilde\alpha_{0i} &= -\frac{i}{\sqrt{2}}\Bigl(\frac{\partial}{\partial x^i} + E_{ki}\frac{\partial}{\partial \tilde x_k} \Bigr) \equiv -\frac{i}{\sqrt{2}} \bar D_i \ .
}
We have thus defined derivatives that will play an important role later
\eqn{
\label{eqderivatives}
	D_i &= \partial_i - E_{ik}\tilde\partial^k \,,\qquad D^i \equiv G^{ij}D_j\,,
	 \\[0.4ex]
	\bar D_i &= \partial_i + E_{ki}\tilde\partial^k \,,\qquad \bar D^i \equiv G^{ij}\bar D_j \ .
}
It turns out that the Virasoro operators with zero mode number 
are given by  
\eqn{
	L_0 &= \frac{1}{2} \alpha_0^i G_{ij}\alpha_0^j + N - 1\,, \\
	\bar L_0 &= \frac{1}{2} \bar\alpha_0^i G_{ij}\bar\alpha_0^j + \bar N - 1 \ ,
}
where $N$ and $\bar N$ are number operators counting the excitations.
There is a constraint in closed string theory which matches the levels of the right and the left moving excitations in any state. It requires that $L_0-\bar L_0=0$. Using the derivatives defined above we can express $L_0 - \bar L_0$ as:
\eqn{
	L_0 - \bar L_0 &= N - \bar N - \frac{1}{4} (D^i G_{ij}D^j - \bar D^i G_{ij}\bar D^j)
		= 	N - \bar N - \frac{1}{4} (D^i D_i - \bar D^i \bar D_i) \,.
}

\begin{Ex}
Show that 
\eqn{
	\frac{1}{2}(D^iD_i - \bar D^i\bar D_i) = -2 \partial_i\tilde\partial^i\ .
}
\end{Ex}

The constraint $L_0-\bar L_0 = 0$ can now be expressed as a constraint on the  number operators in the following way:
\eqn{
\label{eqconstraint}
	N-\bar N &= -\partial_i\tilde\partial^i \equiv -\partial\cdot\tilde\partial\ .
}
The familiar massless fields with $N=\bar N=0$ have the form
\eqn{
	&\sum_{p,w} \; e_{ij}(p, w)\alpha_{-1}^i\bar\alpha_{-1}^jc_1\bar c_1|p,w\rangle, \\
	&\sum_{p,w} \; d(p,w)(c_1c_{-1}-\bar c_1\bar c_{-1})|p,w\rangle \ ,
}
with momentum space wavefunctions $e_{ij}(p,w)$
and $d(p,w)$.  Here the matter and ghost oscillators act on a vacuum $|p,w\rangle$ with momentum $p$ and winding $w$.
On account of (\ref{eqconstraint})  we must require that the fields $e_{ij}(x, \tilde x)$ and $d(x, \tilde x)$ satisfy the constraint
\eqn{
	\partial\cdot\tilde\partial\; e_{ij}(x, \tilde x) = \partial\cdot\tilde\partial \;d(x, \tilde x) = 0 \ .
}
This constraint is a very important ingredient which any string theory and any double field theory has to satisfy.

\subsection{$O(D,D)$ transformations}

It is important to understand the invariance of the physics under background transformations. In particular, $O(D,D)$ transformations play a prominent role in our case. In order to study them we start with the Hamiltonian, which can be constructed from the Hamiltonian density $\underline H$ in (\ref{Hamiltonian}). One can show that 
\eqn{
\label{HamiltonianInt}
	H = \int_0^{2\pi} d\sigma \underline H
		= \frac{1}{2}Z^t {\cal H}(E)Z + N + \bar N + \dots
}
where the dots indicate terms irrelevant to the discussion and 
 $$Z = \begin{pmatrix}w^i\\p_i\end{pmatrix}\,,$$ is a $2D$ column vector consisting of integer winding and momentum quantum numbers.
The $L_0-\bar L_0 =0$ condition (\ref{eqconstraint}) on the spectrum gives
$N- \bar N = p_i w^i$, or equivalently, 
\eqn{\label{Zcon}
	N - \bar N = \frac{1}{2} Z^t\eta Z\ ,}
	where $\eta$ is the matrix defined in (\ref{etaHeta}).
Consider now a reshuffling of the quantum numbers 
$$Z = h^t Z'\,,$$ 
with some $2D\times 2D$ invertible matrix $h$
with integer entries ($h^{-1}$ should also have integer entries). Under such a transformation the physics should not change, and in particular the constraint (\ref{Zcon}) should be unchanged. For this it is then necessary that
\eqn{
	{Z'}^t\eta Z' = Z^t\eta Z = {Z'}^th\eta h^tZ'\ ,
}
which requires
\eqn{
\label{hetaht}
	h\eta h^t = \eta\ .
}

\begin{Ex}
Show that (\ref{hetaht}) implies
\eqn{
\label{htetah}
	h^t\eta h = \eta\ ,
}
\end{Ex}

The $h$ matrices generate the $O(D,D)$ group. We write
\eqn{
	 h = \begin{pmatrix}a & b \\c & d\end{pmatrix}\in O(D,D) 
\,,} 
where $a,b,c$ and $d$ are $D\times D$-matrices.  The
conditions on $a, b, c$, and $d$ following from (\ref{htetah})
are
\eqn{ a^t c + c^t a =  b^t d + d^t b = 0 \,, ~~ a^t d + c^t b = 1 \,. }
The conditions that follow from (\ref{hetaht}) are not independent but they are useful to have
\eqn{ a b^t + ba^t  =  c d^t + d c^t = 0 \,, ~~ a d^t + b c^t = 1 \,. }

\begin{Ex}
Show  that
\eqn{
	h^{-1} &= \begin{pmatrix} d^t & b^t \\ c^t & a^t \end{pmatrix}\ .
}
\end{Ex}

 More is still needed in order to ensure the invariance of the spectrum. The energy, or Hamiltonian must not change. This requires
a change of the background field $E$: the shuffled quantum numbers
are associated to a background field $E'$.  From (\ref{HamiltonianInt}) we demand
\eqn{
	Z^t\HH(E) Z \ = \ {Z'}^t \HH(E') Z' \,.
}
We thus have 
\eqn{
	{Z'}^t h\HH(E) h^t Z' = {Z'}^t \HH(E') Z'\ .
}
We therefore learn that
\eqn{
\label{eq21}
	\HH(E') = h\HH(E) h^t\ .
}
Using the indices introduced in (\ref{oddnotation})  we associate with
$h$ the transformation of coordinates 
\eqn{\label{xtrans}
{X'}^M  =  h^M{}_N \, X^N        \,, ~~~ X \equiv  \begin{pmatrix} \tilde x \\ x 
\end{pmatrix}\,.
}
and then (\ref{eq21}) becomes
\eqn{\label{eq21b}
{\cal H}^{MN} (E') \ = \ h^M{}_P\, h^N{}_Q\, {\cal H}^{PQ}(E) \,.
}

Given that ${\cal H}$ is a rather complicated function of the metric $G$
and the field $B$ associated with $E = G+ B$, it is not obvious that there is a transformation of $E$ that induces the covariant
transformation (\ref{eq21}) (or (\ref{eq21b})) of
${\cal H}$. 
 The transformation of  $E$ in fact exists and is given by:
\eqn{
\label{eq22}
	E' = h(E) = (aE+b)(cE+d)^{-1} \equiv
	\pmat{ a & b\\ c & d } E \ .
}
This is actually a well known transformation which appears often in string theory. It looks like a modular transformation.
The fields $G$ and $B$ in $E$  have  much more complicated transformation laws. This is an indication 
that $E$ is a good variable to formulate our theories.

\begin{Ex}
Show that
\eqn{
\label{eq23}
	{E'}^t = \begin{pmatrix} \phantom{-}a & -b\\-c & \phantom{-}d\end{pmatrix} E^t.
}
\end{Ex}

In order to show that (\ref{eq21}) holds, we first consider the possibility that $E$ is created from the identity background $I$ by the action of $h$. Is it possible to create any such background from the identity~? If so, then this would be a very convenient insight. Let us assume it is true for the moment and assign a transformation $h_E$ to any $E$, so that
\eqn{
\label{eq23b}
	E=h_E(I)\ .
}
To see that $h_E\in O(D,D)$ really does exist we re-write the field $G$ in $E=G+B$. Since $G$ is symmetric it can be written as $G=AA^t$, where $A$ appears like a vielbein.
Using now $A$ and $B$ in the explicit expression for $h_E$ we find that
\eqn{
	h_E=\pmat{ A & B(A^t)^{-1} \\ 0 & (A^t)^{-1} }\,.
}
It is easy to check that $h_E$ is indeed an element of $O(D,D)$. In order to see that it satisfies (\ref{eq23b}) we compute
\eqn{
	h_E(I) = (AI + B(A^t)^{-1})(0\cdot I + (A^t)^{-1})^{-1} = (A+B(A^t)^{-1})A^t = AA^t + B = E \ .
}
This indeed shows that any background $E$ can be created from the identity background by the transformation that we have explicitly constructed.

The transformation  $h_E$ is ambiguous since it is always possible to replace $h_E$ by $h_E \cdot g$ where
$g(I) = I$. These $g$ are elements of $O(D,D)$, and in fact they form a subgroup.

\begin{Ex}
Show that the elements $g$ that satisfy $g(I)=I$ form an $O(D)\times O(D)$ subgroup of $O(D,D)$ and $g^tg=gg^t=I$.
\end{Ex}

With these preparations we can now focus again on (\ref{eq21}) and show that ${\cal H}$ transforms in the right way.
For the construction of $h_E$ we have split the metric $G$ into a product of $A$ and $A^t$, so that only $A$ entered in $h_E$. In order to find 
a matrix with $G$  it is natural to consider the product $h_Eh_E^t$
which does not have the ambiguity of exercise~6. This can be calculated in a straightforward manner:
\eqn{
	h_Eh_E^t = \pmat{A & B (A^t)^{-1} \\ 0 & (A^t)^{-1}} \pmat{A^t & 0 \\-A^{-1}B & A^{-1}} 
	= \pmat{G-BG^{-1}B & BG^{-1} \\ -G^{-1}B & G^{-1}}
	= {\cal H}(E).
}
Suppose now $E'$ is a transformation of $E$ by $h$, i.e.\ $E'=h(E)
= h h_E (I)$. We also have  $E'=h_{E'} (I)$. 
We thus  see that $h_{E'} = h h_E g$, up to the ambiguous $O(D,D)$ subgroup formed by $g$.
Now we can put all this together to compute
\eqn{
	{\cal H}(E') = h_{E'}h_{E'}^t  = hh_Eg(hh_Eg)^t = hh_Eh_E^th^t = h{\cal H}(E) h^t \ .
}
This proves (\ref{eq21}).

Our aim is to show that it is natural to replace the standard notation in string theory based on $G$ and $B$ by $E$ and ${\cal H}$, and in fact we will later re-write the Einstein action completely in terms of ${\cal H}$. In order to arrive there we still need a little more formalism.

First we need to understand how $G$ and $G'$ are related. This relation is not immediately visible. We claim that
\eqn{
\label{eq26a}
	(d+cE)^tG'(d+cE)=G\ .
}
This expression involves $E$ but neither $a$ nor $b$ (from $h$) enter. It looks like a transformation law for tensors, but it is in fact a bit more complicated, since we have $E$-dependent matrices. In the end this will lead to a new kind of indices which are characterized by the fact that they transform like (\ref{eq26a}).

The metric $G$ has the peculiar property that in addition it also satisfies
\eqn{
\label{eq26b}
	(d-cE^t)^tG'(d-cE^t)=G\ .
}
This has some deeper meaning, as we will see.

\begin{Ex}
Prove that
\eqn{
\label{eq26}
	(d+cE)^tG'(d+cE) &=G\,, \\
	(d-cE^t)^tG'(d-cE^t)&=G \ .
}
Hint: Write $G'=\frac{1}{2}(E'+{E'}^t)$ and use (\ref{eq22}) for the first line.  Write $G'=\frac{1}{2}((E')^t+({E'}^t)^t)$  and use (\ref{eq23}) for the second line.
\end{Ex}

In order to sharpen notation let us introduce the matrices
\eqn{
	M &\equiv (d-cE^t)^t,\\
	\bar M &\equiv (d+cE)^t\ .
}
With this abbreviation (\ref{eq26}) becomes
\eqn{
\label{Gtrans}
	G &= \bar M G' \bar M^t\ ,\\
	G &= M G'M^t\ .
}
It is instructive to write these equations in index notation. These are in fact examples of $O(D,D)$ ``tensors,'' which transform in the following way:
\eqn{
	G_{\bar i\bar j} &= {\bar M_{\bar i}\,}^{\bar p} {\bar M_{\bar j}\,}^{\bar q} G'_{\bar p\bar q}\ ,\\
	G_{ij} &= {M_i\,}^p {M_j\,}^q G'_{pq}\ .
}
Note that we have used two kinds of indices for the same object $G$. It is possible to describe $G$ either with barred indices $G_{\bar i\bar j}$ or with unbarred indices $G_{ij}$. Each type of indices comes with a different transformation law, but still they describe the same transformation.

Previously we found  indices $M, N$ that are used for
$O(D,D)$ tensors. Now we found other indices for which $O(D,D)$ transformations are generated by matrices $M$ and $\bar M$.
Thus we want to understand how these two index manipulations are related to each other.
Consider some 
object with components
$$ \Theta^M = \pmat{\tilde\theta_i\\\theta^i}\,.$$
We call such an object a ``fundamental of $O(D,D)$'' if $\Theta'=h\Theta$, or in components
\eqn{
	\pmat{\tilde\theta'\\\theta'} = \pmat{a&b\\c&d}\pmat{\tilde\theta\\\theta} \ ,
}
and we say it transforms in the fundamental representation of $O(D,D)$. Now let us define two more objects
\eqn{\label{ybary}
	Y_i &\equiv -\tilde\theta_i + E_{ij}\theta^j\,, \\
	\bar Y_i &\equiv \phantom{-}\tilde\theta_i + E_{ji}\theta^j \,,
}
using the $\theta$'s and the $E$. These objects will not transform just with $h$, since they now depend on $E$. Still, they have a simple transformation law, involving the $M$'s:
\eqn{
\label{eq32}
	Y_i &= {M_i\,}^j Y'_j\ ,\\
	\bar Y_{\bar i} &= {\bar M_{\bar i}\,}^{\bar j}\bar Y_{\bar j}'\ . 
}
Thus the above construction tells us how to move from an object $\Theta$ which transforms with $h$ to an object $Y$ which transforms with $M$.

\begin{Ex}
Prove the first line of (\ref{eq32}). For this use, and prove, the identity
\eqn{
\label{eq33}
	b^t-Ea^t=-ME' \ .
}
\end{Ex}

This has a useful application.
Consider a fundamental object
$$X^M = \pmat{\tilde x_i\\x^i}\,.$$ 
The associated partial derivative is
\eqn{
\partial_M = \pmat{\tilde\partial^i\\\partial_i}\, ~~\to ~~ 
\partial^M\equiv \eta^{MN}\partial_N = \pmat{\partial_i\\\tilde\partial^i}\,.
}
The derivative $\partial^M$ is also in the fundamental representation. From this it is now possible to calculate 
\eqn{
\partial^M\partial_M = 2\partial_i\tilde\partial^i=0\,. } 
This is recognized as the constraint (\ref{eqconstraint}).
In the same way as we have constructed the objects $Y$ and $\bar Y$ in (\ref{ybary}) above, we can construct derivatives transforming under the action of $M$. When we do that we find that the natural objects to write  are
\eqn{
	-\partial_i + E_{ij}\tilde\partial^j &= -D_i\,,\\
	\partial_i+E_{ji}\tilde\partial^j &= \phantom{-}\bar D_i\ ,
}
which are exactly the derivatives in (\ref{eqderivatives}). So we see, those derivatives we find in string theory are in fact $O(D,D)$-derivatives and
transform covariantly under $O(D,D)$:
\eqn{
	D_i &= {M_i\,}^jD'_j \,,\\
	\bar D_i &= {\bar M_i\,}^j\bar D'_j\,.
}

The last object whose transformation properties we have to understand better is that for the variation $\delta E$ of the background field.
We know already that $E'=h(E)$, which is a complicated expression when written out. While $E$ does not transform as a tensor, its 
variation does.
 We find
\eqn{
	E' + \delta E' &= h(E+\delta E) \\
	&= (a(E+\delta E)+b)(c(E+\delta E)+d)^{-1} \\
	&= (aE + b + a\delta E)(cE +d + c\delta E)^{-1} \\
	&= E' + a \delta E (cE+d)^{-1} - E'c\delta E(cE+d)^{-1} \ ,
}
where we used $(A+\varepsilon)^{-1}=A^{-1}-A^{-1}\varepsilon A^{-1} + O(\varepsilon^2)$ in the last step.  From this we get
\eqn{\label{vmvg}
	\delta E' =  (a- E'c) \delta E (cE+d)^{-1} =\ (a- E'c) \delta E 
	({{\bar M}^t})^{-1}  \  \ .
}
The last hurdle is a bit of manipulation.  

\begin{Ex}
Prove that  $a- E'c= M^{-1}$.  For this check that $M (a- E'c) = 1$
by explicit multiplication, using the results of Exercise 8.
\end{Ex}

From (\ref{vmvg}) and the result of the above exercise  one reads off the transformation law
\eqn{
\label{eq35}
	\delta E = M\delta E' \bar M^t \ .
}
We see that $E$ has one unbarred index and one barred index:
\eqn{
	\delta E_{i\bar j} = {M_i}^p  \bar M_{\bar j}^{\;\bar q} \delta E'_{p\bar q}\ .
}

We have set up a consistent formalism and have understood the transformation laws of the fundamental objects in our theory. We can use this in order to construct actions.

\section{Double Field Theory Actions}

For the construction of actions using the previously developed formalism we start with a background field $E_{i\bar j}$ and small fluctuations $e_{i\bar j}(x,\tilde x)$. This should be thought of as a background configuration which contains a gravitational background as well as a background Kalb-Ramond field. In addition we include a dilaton $d(x,\tilde x)$.

\subsection{The quadratic action}

First we focus on the quadratic part of the action, given by
\eqn{
\label{eq41}
	S^{(2)} &= \int dxd\tilde x\Bigl[
		\frac{1}{4} e^{i\bar j}\Box e_{i\bar j} + \frac{1}{4}(\bar D^{\bar j} e_{i\bar j})^2 + \frac{1}{4}(D^i e_{i\bar j})^2
			- 2d D^i\bar D^{\bar j} e_{i\bar j} - 4d\Box d
	\,\Bigr]\,,
} 
where indices are raised by the background metric $G^{ij}$ (or $G^{\bar i\bar j}$) and the box operator is given by $\Box = D^iD_i = \bar D^{\bar i}\bar D_{\bar i}$ with constraint $D^2-\bar D^2=0$. This constraint is equivalent to $\partial\cdot\tilde\partial=0$ and must be satisfied by all fields and gauge parameters.

Under $O(D,D)$-transformations the objects under the integral will transform with $M$ or $\bar M$. Note that $M$ and $\bar M$ depend on the background field $E$ and not on the fluctuations~$e_{i\bar j}$. This implies that derivatives will not act on $M$ or $\bar M$. So this action is manifestly $O(D,D)$-invariant.

This action must have gauge symmetries, which must include those found in general relativity. In fact gauge symmetry fixes the relative values of the coefficients of the terms in the action.

\begin{Ex}
Prove that the following are gauge invariances of $S^{(2)}$:
\eqn{\label{ginvs2}
	\delta e_{i\bar j} = \bar D_{\bar j} \lambda_i\ , &\qquad \delta e_{i\bar j} = D_i\bar\lambda_{\bar j}\ ,\\
	\delta d = -\frac{1}{4} D^i \lambda_i\ , &\qquad \delta d = -\frac{1}{4}\bar D^{\bar i}\bar\lambda_{\bar i}\ .
}
\end{Ex}

In order to get a better feeling for this action we write it out more explicitly, simplifying it by setting the background Kalb-Ramond field $B_{ij}$ to zero, keeping only the fluctuations $e_{i\bar j} = h_{ij} + b_{ij}$ around the metric $G_{ij}$.
The action becomes
\eqn{
\label{eqquadaction}
	S^{(2)} = \int dxd\tilde x \,\Bigl[\,
		& \frac{1}{4}h^{ij}\partial^2h_{ij} + \frac{1}{2}(\partial^i h_{ij})^2
		- 2d \partial^i\partial^j h_{ij} - 4d\partial^2 d \\
		+&\frac{1}{4}h^{ij}\tilde\partial^2 h_{ij} + \frac{1}{2}(\tilde\partial h_{ij})^2 + 2d\tilde\partial^i\tilde\partial^j h_{ij}
		- 4d\tilde\partial^2 d \\
		+&\frac{1}{4}b^{ij}\partial^2b_{ij} + \frac{1}{2}(\partial^j b_{ij})^2 \\
		+&\frac{1}{4}b^{ij}\tilde\partial^2 b_{ij} + \frac{1}{2}(\tilde \partial^j b_{ij})^2 \\
		+ &(\partial_k h^{ik})(\tilde\partial^j b_{ij}) + (\tilde\partial^k h_{ik})(\partial_j b^{ij}) 
		- 4d\partial^i\tilde\partial^j b_{ij}
	~\Bigr]\,.
}
The first line contains the graviton and dilaton in the same way as one would get from the standard action. The second line is almost identical to the first line but contains dual derivatives. This is to be expected since the whole action should be $O(D,D)$-invariant. The third line contains the contributions of the Kalb-Ramond field strength, while the fourth line again complements it with terms involving dual derivatives. The last line contains terms with mixed derivatives. These terms have no counterpart in conventional field theory actions.

The symmetries of this action are conveniently described in terms of redefined gauge parameters
\eqn{
	\epsilon_i &= \frac{1}{2}(\lambda_i+\bar\lambda_i)\qquad\tilde\epsilon_i = \frac{1}{2}(\lambda_i-\bar\lambda_i)\ .
}
Using these the gauge symmetries (\ref{ginvs2}) are found to be
\eqn{\label{var1}
	\delta h_{ij} = \partial_i\epsilon_j + \partial_j\epsilon_i 
		&\qquad \tilde\delta h_{ij} =  \tilde\partial_i\tilde\epsilon_j + \tilde\partial_j\tilde\epsilon_i\\
	\delta b_{ij} = -(\tilde\partial_i\epsilon_j - \tilde\partial_j\epsilon_i)
		&\qquad \tilde\delta b_{ij} = -(\partial_i\tilde\epsilon_j-\partial_j\tilde\epsilon_i) \\
	\delta d = -\frac{1}{2}\partial\cdot\epsilon
		&\qquad \tilde\delta d =  \frac{1}{2}\tilde\partial\cdot\tilde\epsilon\ .
}
To appreciate the novel aspects of the above, consider the familiar
linearized gauge symmetries of the low energy (non-double) action (\ref{lowenergyaction}):
\eqn{\label{var2}
	\delta h_{ij} = \partial_i\epsilon_j + \partial_j\epsilon_i 
		&\qquad \tilde\delta h_{ij} =  ~~0 \,,\\
	\delta b_{ij} = ~~0\,,~~~ 
		&\qquad \tilde\delta b_{ij} = -(\partial_i\tilde\epsilon_j-\partial_j\tilde\epsilon_i) \\
	\delta d = -\frac{1}{2}\partial\cdot\epsilon
		&\qquad \tilde\delta d =  ~~ 0 \,  .
}
In (\ref{var2}) we have two columns.  The left one corresponds to the
symmetry of diffeomorphisms, with gauge parameter $\epsilon_i$.  
The gravity fluctuation transforms, the $b$ field does not, and the dilaton
$d$ transforms as a scalar density. The conventional scalar dilaton
$\Phi$ is given by $\Phi \equiv d+\frac{1}{4}{h^i}_i$ and is gauge invariant. 
In the double field theory case (\ref{var1}) the $b$ field transforms using the tilde derivatives to form the required antisymmetric right-hand side. 

In the second column of (\ref{var2}) the gauge parameter $\tilde\epsilon_i$
generates the $b$ field transformations.  No other field transforms under it.
But in the corresponding column of (\ref{var1}) we see $h$ transforming
under what we could call dual diffeomorphisms and $d$ transfoming as a dual density.    The combination
$\tilde\Phi \equiv d-\frac{1}{4}{h^i}_i$ is invariant under the $\tilde \epsilon$
symmetry.   Since $\tilde \Phi$ is not invariant under the $\epsilon$ transformation nor is $\Phi$ invariant under $\tilde\epsilon$ transformations there is no dilaton that is a scalar under {\em both}
diffeomorphisms and dual diffeomorphisms.

\subsection{The cubic action}

For going beyond the free theory cubic terms should be added to the action. Indeed there are cubic terms which are $O(D,D)$-invariant and can be added consistently to the action. This results in fact in a nonlinear extension of the gauge invariance.

For simplicity we focus only on a few possible terms in the cubic part
$S^{(3)}$ of the action and refer to the literature for complete details:
\eqn{
\label{eqcubicaction}
	S^{(3)} = \int dxd\tilde x
		\;\frac{1}{4}&e_{ij}\Bigl((D^ie_{kl})(\bar D^je^{kl}) - D^ie_{kl}\bar D^le^{kj} - D^ke^{il}\bar D^je_{kl}\Bigr) \\[1.0ex]
		 &+de^2\text{ terms}
		+ d^2e\text{ terms}
		+ d^3\text{ terms}\ .
}
The nonlinear extension of the gauge symmetry can be seen from the variation of $e$, which is given by
\eqn{
\label{eqcgauge}
	\delta_\lambda e_{i\bar j} &= \bar D_{\bar j}\lambda_i + \frac{1}{2}\Bigl[
		(D_i\lambda^k)e_{k\bar j} - (D^k\lambda_i)e_{k\bar l} + \lambda_k D^ke_{i\bar j}
	\Bigr] \ .
}
While the construction up to cubic order has been completed, higher orders may be very nontrivial. It may even happen that higher orders do not  exist as long as one restricts oneself to a formulation involving only the massless fields $e_{ij}$ and $d$.

\medskip
We have stressed that all fields and gauge parameters must satisfy the constraint that they are annihilated by $\partial\cdot\tilde\partial$. This was enough for the quadratic action and in fact for the cubic action. 
But even for the gauge transformations (\ref{eqcgauge}) there is an important subtlety.   It is not true that $\partial\cdot\tilde\partial$ annihilates a product of two fields, even if each field is annihilated individually.
Thus the terms in brackets in  (\ref{eqcgauge}) do not satisfy the contraint; they should since they represent a variation of the constrained field 
$e_{i\bar j}$.  Thus one must include for those terms in brackets a projector to the space of functions that satisfy the constraint. Such projectors are not needed in the cubic action (the integration does the
projection automatically) but they complicate matters considerably when trying to construct the quartic terms of the action.

To be able to proceed more simply we impose a stronger constraint. 
We simply demand that the operator $\partial\cdot\tilde\partial$ annihilates all fields {\em and}  all products of fields.

Let $A_i(x,\tilde x)$ be fields or gauge parameters which are annihilated by $\partial_M\partial^M$. When we require now that all products $A_iA_j$ be also killed by $\partial_M\partial^M$ this leads to the condition
\eqn{
\label{eq50}
	\partial_MA_i\partial^MA_j=0\,, ~~~ \forall \, i, j \,.
}
We may call this the ``strong'' $O(D,D)$  constraint.

In fact this is a {\em very} strong constraint, and while it makes the calculations easier it makes us lose much physics. It turns out that this strong constraint makes the theory independent of the dual coordinates in the following sense:

\begin{Thm}
For a set of fields $A_i(x,\tilde x)$ that satisfies (\ref{eq50}) there is a duality frame $(\tilde x'_i, {x'}^i)$ in which the fields do not depend on $\tilde x_i'$.
\end{Thm}

Even if it is always possible to find such a frame, we need not specify it explicitly, i.e.\ we need not break $O(D,D)$ invariance. The constraint (\ref{eq50}) is indeed $O(D,D)$ invariant. Hence we are in a situation where we can formulate a theory using dual coordinates in the action, keeping the full $O(D,D)$ invariance, while physically only half of the coordinates matter.

\section{Courant brackets}

In a theory with a metric $g_{ij}(x)$ and a Kalb-Ramond field $b_{ij}(x)$ the  diffeomorphisms are generated by vector fields $V^i(x)$ and Kalb-Ramond gauge transformations are generated by one-forms $\xi_i(x)$. These are formally added and thus written as $V+\xi \in T(M)\oplus T^*(M)$, where $V\in T(M)$ and $\xi\in T^*(M)$ are elements of the tangent bundle and the cotangent bundle, respectively. We can formulate the gauge transformations in a geometric language
\eqn{
\label{eq51}
	\delta_{{}_{V+\xi}} \,g &= {\cal L}_V g\,, \\[0.3ex]
	\delta_{{}_{V+\xi}}\, b &= {\cal L}_Vb + d\xi\ ,
}
where ${\cal L}_V$ is the Lie derivative along the vector field $V$.
Recall that when acting on forms the Lie derivative is
\eqn{
	{\cal L}_V=\iota_V d + d \iota_V\ ,
}
where $\iota_V$ is contraction with $V$. It follows that the Lie derivative and the exterior derivative commute,
\eqn{
\label{eq53}
	{\cal L}_Vd = d{\cal L}_V\ .
}
Acting on the metric the Lie derivatives gives 
\eqn{
	({\cal L}_V\,g)_{ij} &= (\partial_iV^k)g_{kj} + \partial_jV^kg_{ik} + V^k\partial_kg_{ij}\ .
} 
Lie derivatives satisfy interesting algebraic relations:
\eqn{
	[{\cal L}_X, {\cal L}_Y] &= {\cal L}_{[X,Y]} \,,\\
	[{\cal L}_X, \iota_Y] &= \iota_{[X,Y]} \ .
}
The left hand sides are commutators of operators and on the
right-hand side  we find brackets of vector fields, defined as $[V_1, V_2]^k =V_1^p\partial_pV_2^k - (1\leftrightarrow 2)$. 

\subsection{Motivating the Courant bracket}

Suppose one has a theory of a metric and an antisymmetric tensor field and one has derived the transformation laws (\ref{eq51}), how can one determine the gauge algebra?
First we  compute the algebra of gauge transformations on the metric $g$ by evaluating the bracket
\eqn{
	[\delta_{V_2+\xi_2}, \delta_{V_1+\xi_1}] g 
		&= {\cal L}_{V_1}{\cal L}_{V_2}g - (1 \leftrightarrow 2)
		= {\cal L}_{[V_1,V_2]} g \ .
}
On the Kalb-Ramond field $b$ the computation is a little less trivial. We find
\eqn{
	[\delta_{V_2+\xi_2}, \delta_{V_1+\xi_1}] b
	&= {\cal L}_{V_1}({\cal L}_{V_2}b+d\xi_2) - (1 \leftrightarrow 2)
	= {\cal L}_{[V_1,V_2]} b + d({\cal L}_{V_1}\xi_2 - {\cal L}_{V_2}\xi_1)\ .
}
When we compare this with (\ref{eq51}) we conclude that acting on the  fields
\eqn{
	[\delta_{V_2+\xi_2}, \delta_{V_1+\xi_1}] &= \delta_{[V_1,V_2] + {\cal L}_{V_1}\xi_2 - {\cal L}_{V_2}\xi_1} \ .
}
This last expression defines a ``bracket'' on $T(M)\oplus T^*(M)$:
\eqn{\label{bracketnaive}
	[V_1+\xi_1, V_2+\xi_2] = [V_1, V_2] + {\cal L}_{V_1}\xi_2 - {\cal L}_{V_2}\xi_1\ .
}
The first term on the right-hand side is a vector field, the last two give
a one-form. One may ask now if this bracket is a Lie bracket. It is because it is antisymmetric and the Jacobi identity is satisfied (as a calculation shows).

There is, however, an ambiguity in the one-form because this one-form 
appears in the gauge transformation acted by the exterior derivative. 
Indeed,
$$\delta_{V+\xi}b = {\cal L}_Vb + d\xi = {\cal L}_{V+(\xi+d\sigma)}\, b\,.$$
 Thus the one-form $\xi$ is ambiguous up to an exact term $d\sigma$. 
This ambiguity also is present in (\ref{bracketnaive}). To see this we calculate the exterior derivative of the form on the right-hand side
\eqn{
	d({\cal L}_{V_1}\xi_2 - {\cal L}_{V_2}\xi_1)
	&= d(\underline{d \iota_{V_1}\xi_2} + \iota_{V_1}d\xi_2 - (1\leftrightarrow 2))
}
The underlined term is killed by the action of $d$, so without loss of generality we may change the coefficient in front of it. We will do so by replacing it with $1-\frac{\beta}{2}$:
\eqn{
	d({\cal L}_{V_1}\xi_2 - {\cal L}_{V_2}\xi_1)
	&= d\Bigl({\cal L}_{V_1}\xi_2 - {\cal L}_{V_2}\xi_1 - \frac{1}{2}\beta d(\iota_{V_1}\xi_2 - \iota_{V_2}\xi_1)\Bigr) \ .
}
This ambiguity should be reflected in our definition of the bracket. So we replace (\ref{bracketnaive})~by 
\eqn{
\label{eq60}
	[V_1+\xi_1, V_2+\xi_2]_\beta &= [V_1, V_2] + {\cal L}_{V_1}\xi_2 - {\cal L}_{V_2}\xi_1 - \frac{1}{2}\beta d(\iota_{V_1}\xi_2 - \iota_{V_2}\xi_1) \ .
}
One complication with this bracket is that is does not satisfy a Jacobi identity as long $\beta$ does not vanish.
Does it make sense to consider brackets with $\beta\ne0$ at all~? Yes it
does!   One can show that, with
$Z_i = V_i+\xi_i$, $i=1,2,3$, the ``Jacobiator" takes the form
\eqn{
	[Z_1, [Z_2, Z_3]] + \text{cyclic} = dN(Z_1, Z_2, Z_3)\ .
}
The right hand side is not zero but an exact 1-form.  Since exact one-forms do not generate gauge transformations, the failure of the Jacobi 
identity does not cause inconsistency.

This bracket is not a new invention, but it has been considered before by T.~Courant in 1990. He had reasons to fix $\beta=1$ and therefore defined a bracket called the {\it Courant bracket} as 
\eqn{
\label{eq61}
	[V_1+\xi_1, V_2+\xi_2]_{\beta=1} &= [V_1, V_2] + {\cal L}_{V_1}\xi_2 - {\cal L}_{V_2}\xi_1 - \frac{1}{2} d(\iota_{V_1}\xi_2 - \iota_{V_2}\xi_1) \ .
}

In fact for $\beta=1$ there is an extra automorphism of the bracket, called $B$-transformation. This is what makes it  interesting from a  mathematical point of view. Given a closed 2-form $B$ with $dB=0$, the $B$-transformation acts on a pair $(X, \xi)$ of gauge parameters as follows, 
\eqn{B-\hbox{transformation:}~~~
	X+\xi \mapsto X+(\xi+\iota_X B)\ .
}
So this map has the effect that it changes the 1-form. If $B$-transformations are an automorphism of the bracket one must have:
\eqn{
	[X+\xi+\iota_XB, Y+\eta+\iota_YB] &= [X+\xi, Y+\eta] + \iota_{[X,Y]} B \ .
}

\begin{Ex}
Show that the existence of this automorphism selects $\beta=1$ in (\ref{eq60}), thus giving (\ref{eq61}).
\end{Ex}

The reason why automorphisms like the $B$-transformation are interesting for us is that they tell us something about the symmetries of a theory. Consider a manifold with some metric $g$. We say that some vector field $V$ is an isometry (and therefore generates a symmetry of the metric) if the Lie derivative ${\cal L}_Vg$ vanishes. If we have an anti-symmetric field $b$  on a manifold, one is tempted to demand that symmetries 
correspond to vector fields for which the Lie derivative of $b$ vanishes. In fact this is too restrictive. Instead it is reasonable to demand that ${\cal L}_Vb$ vanishes up to some exact form, since any such change of $b$ can be undone by a $b$-field gauge transformation.
Therefore, $V+\xi\in TM\oplus T^*M$ is a symmetry of $b$ if
\eqn{
	{\cal L}_V b = d\xi\ .
}
Consider a 2-form $B$ with $dB=0$. Imagine changing $b$ by adding
$B$ to it.  What are the symmetries of the new $b+B$ field? We claim that the $B$-transform of $V+\xi$ is a symmetry of $b+B$,
\eqn{
	{\cal L}_V(b+B) 
	= d(\xi+\iota_VB)\ .
}
It is straightforward to verify this by explicit calculation.
From this we see that $B$-trans\-for\-mations of $b$ do not change the symmetries of the theory. Thus it is reasonable to  promote $B$-trans\-for\-mations to automorphisms of the bracket, thus selecting the Courant-bracket.

\subsection{Algebra of Gauge Transformations: from Courant brackets to 
$C$ brackets}

In order to determine the algebra of gauge transformations we switch to a more uniform notation in which we mark all one-forms by tildes while vectors stay undecorated. Hence we consider objects
$$\xi^M=\begin{pmatrix} \tilde\xi_i \\\xi^i\end{pmatrix}\,, $$
denoting gauge parameters in the sum of tangent and cotangent space of the manifold. In an abuse of notation we sometimes write this as 
$$\xi^M=(\xi+\tilde\xi)^M\,.$$ 
The gauge algebra is governed by a $C$-bracket $[\,\cdot\, , \cdot\,]_C$, which is closely related to the Courant-bracket but applies to doubled fields! 
The Courant bracket does not, of course. Consider the $M^\text{th}$ component of such a bracket:
\eqn{
	\Bigl( [\xi_1,\xi_2]_C\Bigr)^M
	&= \xi_{[1}^P\partial_P\xi_{2]}^M - \frac{1}{2}\eta^{MN}\eta_{PQ}\xi^P_{[1}\partial_N\xi_{2]}^Q \\
	&= \xi_{[1} \cdot \partial \xi_{2]}^M - \frac{1}{2}\xi_{P[1} \partial^M\xi_{2]}^P \ ,
}
where the brackets on indices indicate anti-symmetrization. Because of the consistent use of our capitalized indices $M, N, \ldots$, this bracket is
$O(D,D)$ covariant. Note that the second term on the right-hand side involves a contraction of indices and therefore contains the metric $\eta$. In a conventional theory it would be 
unthinkable to include a metric-dependent term in a bracket. In our case
the use of the constant metric $\eta$ causes no complications.

Evaluating this bracket between $\xi_1+\tilde\xi_1$ and $\xi_2+\tilde\xi_2$ displays the appearance of some unusual terms:
\eqn{
\label{eq69}
	[\xi_1+\tilde\xi_1, \xi_2+\tilde\xi_2]_C
	= &[\xi_1,\xi_2] + {\cal L}_{\tilde\xi_1}\xi_2 - {\cal L}_{\tilde\xi_2}\xi_1
		- \frac{1}{2}\tilde d(\tilde \iota_{\tilde\xi_1}\xi_2 - \tilde \iota_{\tilde\xi_2}\xi_1) \\
		+ &[\tilde\xi_1,\tilde\xi_2] + {\cal L}_{\xi_1}\tilde\xi_2 - {\cal L}_{\xi_2}\tilde\xi_1
		- \frac{1}{2} d(\iota_{\xi_1}\tilde\xi_2 - \iota_{\xi_2}\tilde{\xi_1})\ ,
}
where the dual exterior derivatives acting on functions give
objects with a vector (upper) index: $(\tilde d f)^i \equiv \tilde \partial^i f$.
It is unusual to see ${\cal L}_{\tilde\xi_2}\xi_1$, since Lie derivatives are taken with respect to  vector fields and not one-forms. In our case  this alternative is allowed since we have (dual) derivatives with upper indices, so that a contraction with a one-form is possible.
In the same way it is no surprise to see a bracket of one-forms giving
a one-form (an object with a lower index): $[\tilde\xi_1,\tilde\xi_2]_j \equiv \tilde{\xi}_{[1_i}\tilde\partial^i\tilde{\xi_{2]}}_j$. 

If we drop the $\tilde x$-dependence of the $C$-bracket this will set ${\cal L}_{\tilde\xi}\to 0$, $~\tilde d\to 0$ and $[\tilde\xi,\tilde\xi]\to 0$. The 
$C$ bracket reduces to
\eqn{
	[\xi_1+\tilde\xi_1, \xi_2+\tilde\xi_2]_C\Bigr|_{\tilde x\equiv 0} &= [\xi_1,\xi_2] + {\cal L}_{\xi_1}\tilde\xi_2 - {\cal L}_{\xi_2}\tilde\xi_1
		- \frac{1}{2}d (\iota_{\xi_1}\tilde\xi_2 - \iota_{\xi_2}\tilde\xi_1)\ .
}
We recognize the right-hand side as the Courant-bracket (\ref{eq61}). Therefore we can view the $C$-bracket as $O(D,D)$ covariant, double field theory generalization of the Courant-bracket.
It can be shown that the $\beta$-parameter cannot be incorporated into the C bracket while preserving  
$O(D,D)$ covariance. 

\subsection{$B$-transformations}

Having identified the algebraic basis of our theory, we now want to understand what are the $B$-transformations in our setup. Take an element of $O(D,D)$, 
\eqn{
	h = \begin{pmatrix} 1 & b \\ 0 & 1\end{pmatrix}\ , 
}
where $b$ is antisymmetric and constant. Acting with this map on $E$ it is easy to compute the transformation 
\eqn{
	E \mapsto E' &= h(E) = (E+b)(1)^{-1} = E+b\ .
}
From this one can read off that the transformation $h$ has the effect of leaving $G$ untouched while $B$ is mapped to $B+b$. So indeed $h$ is a $B$-transformation.
Now it is straightforward to see the action of this map on the gauge 
parameters $\xi^M$.
Explicit evaluation shows that
\eqn{
	\begin{pmatrix}\tilde\xi\\\xi\end{pmatrix} \mapsto \begin{pmatrix} 1 & b \\ 0 & 1\end{pmatrix}
		\begin{pmatrix}\tilde\xi \\ \xi\end{pmatrix}
		= \begin{pmatrix} \tilde\xi+b\xi \\ \xi \end{pmatrix}\,,
}
so that in components the $B$-transformation is given by
\eqn{\label{eq93}
	\xi^i \mapsto \xi^i\,, \qquad
	\tilde\xi_i\mapsto\tilde\xi_i+b_{ij}\xi^j\,,\qquad
	\tilde\partial^i\mapsto\tilde\partial^i \ .
}
Note that invariance of the dual derivatives  implies that  $B$-transformations leave the constraint ${\partial\over \partial \tilde x} \phi= 0$
appropriate for the Courant bracket unchanged. One sees that
(\ref{eq93}) implies
$$\xi+\tilde\xi\to\xi+\tilde\xi+\iota_\xi b\,,$$ which is exactly the expected result.

We see now how nicely the parts fit together to form a larger picture: from the physics point of view we have arrived at this formulation because we took T-duality seriously and considered it as basic component of our field theory. From the mathematics point of view the $B$-transformations play a fundamental role as automorphisms of the Courant-bracket, and in fact now we see that they are just the counterpart of certain T-duality 
transformations that must be incorporated in an $O(D,D)$ invariant formulation.

\section{Background Independent Action}

We now want to put the various parts together and come to a formulation of a doubled action.
We have written down before the perturbative action for a double field theory in terms of a background $E_{ij}$ and fields $e_{ij}(x,\tilde x)$, depending on both the usual coordinates $x$ and their duals $\tilde x$. We made an explicit distinction between the background field and its fluctuation, very similar to the splitting $g_{ij} = \eta_{ij} + h_{ij}$ in linearized gravity. In the end, however, one is looking for a manifest background independent version of the action which does not rely on this distinction. 

\subsection{Background Independent Formulation}

To stress the point of background independence we introduce the field
\eqn{
\label{Edef}
	{\cal E}_{ij}(X) = E_{ij}+e_{ij}(x,\tilde x)+ O(e^2)\,,
}
which at the linearized level is the sum of $E$ and $e$.
We have seen how $E$ and $e$ behave under T-duality, and there is also a natural way to transform ${\cal E}$. Since $X'=hX$ (recall (\ref{xtrans}))  
we 
expect that ${\cal E}$ transforms like
\eqn{
	\EE'(X') = \left(a \EE(X) + b \right)  \left(c \EE(X) + d \right)^{-1} .
	\label{EEtransf}
}
The dilaton $d$ is expected to be $O(D,D)$ invariant, so its transformation law should be
\eqn{
	d'(X') = d(X).
}
This is the analogue of the scalar field Lorentz transformation in conventional field theory.

All the identities and constructions presented in previous sections above did not make use of any $X$-independence of $E$. Therefore they can be immediately generalized by replacing $E$ with ${\cal E}$, keeping the formal expressions unchanged. For example the derivatives $D_i$ in (\ref{eqderivatives}) can be generalized to curly $\DD_i$, and similarly for the $\bar{\DD}_i$'s; but now they are defined with the full metric $\EE$,
\begin{equation}
\begin{split}
D_i = \partial_i - E_{ik}\tilde{\partial}^k \quad & \longrightarrow \quad \DD \equiv \partial_i - \EE_{ik}(X) \tilde{\partial}^k \,,\\[0.3ex]
\bar{D}_i = \partial_i + E_{ki}\tilde{\partial}^k \quad & \longrightarrow \quad \bar{\DD} \equiv \partial_i + \EE_{ki}(X) \tilde{\partial}^k.
\end{split}
\label{DD}
\end{equation}
These derivatives will now transform with generalized $M$ matrices, that now depend on ${\cal E}(X)$ as
\begin{equation}
\begin{split}
M = \left( d - c E^t \right)^t \quad & \longrightarrow \quad M(X) = \left( d - c \EE^t \right)^t\,, \\
\bar{M} = \left( d + c E^t \right)^t \quad & \longrightarrow \quad \bar{M}(X) = \left( d + c \EE^t \right)^t.
\end{split}
\end{equation}
Indeed, any object will now transform correctly with $M(X)$ and $\bar{M}(X)$ exactly in the way as they transformed with $M$ and $\bar M$ before. This is because the transformations come from (\ref{EEtransf}), and one needs no derivatives to derive them. We will also 
write  ${\cal E}=g+b$ without any reference to a background field, and the generalized metric in (\ref{Hamiltonian2}) becomes
\eqn{
	\HH(\EE) &= \begin{pmatrix} 
		g - b g^{-1} b & b g^{-1} \\
		-g^{-1} b & g^{-1} 
		\end{pmatrix}
	\ .
}
In particular, the metric $g(X)$ itself is an $O(D,D)$ tensor, so from (\ref{Gtrans}) we have
\eqn{
	g(X) &= \bar{M}(X) \, g'(X') \, \bar{M}^t(X)\,, \\[0.5ex]
	g(X) &= M(X) \, g'(X') \, M^t(X)\ .
}
Moreover, the transformation of the Hamiltonian in eq.~(\ref{eq21}) becomes
\begin{equation}
\HH(\EE'(X')) = h \, \HH(\EE(X)) \, h^t \ .
\end{equation}
We can repeat all the steps that gave the transformation law
 (\ref{eq35}) for the variation of ${\cal E}$, this time finding
\begin{equation}
\delta \EE(X) = M(X) \, \delta E'(X') \, \bar{M}^t(X)\ .
\end{equation}
This relation applies to any derivative of ${\cal E}$, thus, for example,
\eqn{\partial_i \EE = M(X) \, \partial_i \EE' \, \bar{M}^t(X)\,, 
~~  \tilde\partial^i \EE = M(X) \, \tilde\partial^i \EE' \, \bar{M}^t(X)\,.}
This also means that the same transformations apply to the calligraphic derivatives of ${\cal E}$:
\eqn{{\cal D}_i \EE = M(X) \, {\cal D}_i \EE' \, \bar{M}^t(X)\,, 
~~  \bar{\cal D}_j \EE = M(X) \, \bar{\cal D}_i \EE' \, \bar{M}^t(X)\,.}
The derivatives above can also be transformed, if desired 
(see (\ref{caldtr}) below). 
Finally, the transformation of the dilaton under gauge transformation is given by
\begin{equation}
\delta d = -\frac{1}{2} \, \partial_M \xi^M + \xi^M \partial_M d \ .
\end{equation}
This implies that
\begin{equation}
\delta e^{-2 d} = \partial_M \left[ \xi^M e^{-2 d} \right]\ ,
\end{equation}
which tells us that $e^{-2d}$ is a density. Therefore it is identified as $\sqrt{-g} \, e^{-2 \phi} = e^{-2d}$.

There is one small complication which appears when one takes multiple derivatives. To understand this, we observe that the derivatives (\ref{DD}) transform covariantly
\begin{equation}
\label{caldtr}
\begin{split}
\DD_i &= M_i^{\phantom{i}j}(X) \, \DD'_j \,,\\[0.3ex]
\bar{\DD}_i &= \bar{M}_i^{\phantom{i}j}(X) \, \bar{\DD}'_j\, .
\end{split}
\end{equation}
Since $M$ is not a constant anymore, multiple derivatives would not
transform correctly. We handle this problem simply by not using higher derivatives in the formulation of our action. We can define $O(D,D)$
covariant derivatives, but they will not be needed here.

\subsection{The $O(D,D)$ Action}

After these preparations we can now present the full background independent $O(D,D)$ action for the fields $\EE$ and $d$. The action is given by
\begin{equation}
\begin{split}
S_{\EE, d} = \int dx \, d \tilde{x} \, e^{-2 d} ~\Bigl[ 
& -\frac{1}{4} \, g^{ik} g^{j\ell} \DD^p \EE_{k \ell} \DD_p \EE_{ij} \\
& + \frac{1}{4} \, g^{k\ell} \left( \DD^j \EE_{ik} \, \DD^i \EE_{j\ell} + \bar{\DD}^j \EE_{ki} \, \bar{\DD}^i \EE_{\ell j} \right) \\
& + \left( \DD^i d \, \bar{\DD}^j \EE_{ij} + \bar{\DD}^i d \, \DD^j \EE_{ji} \right) 
+ 4 \, \DD^i d \, \DD_i d ~\Bigr]\ .
\end{split}
\label{S1}
\end{equation}
Each term is independently $O(D,D)$ invariant, and so is the whole action. This also means, though, that the action is not completely determined by $O(D,D)$ invariance, since the numerical factor in front of each term is arbitrary.
What finally fixes the action is diffeomorphism and Kalb-Ramond gauge invariance. There is a particular combination of the coefficients, so that the theory is consistent and exhibits these expected gauge invariances. Also, one can expand this action and recover to quadratic and cubic part of the action exactly as in (\ref{eq41}) and (\ref{eqcubicaction}).
Moreover, taking $\tilde{\partial} = 0$, $S_{\EE, d}$ reduces to an action that is identical to the standard Einstein action plus antisymmetric field plus dilaton, 
when $\sqrt{-g} \, e^{-2 \phi} = e^{-2 d}$. So all this is consistent and fixes the action uniquely.

This action is invariant under the following gauge transformations
\begin{equation}
\begin{split}
\delta_\xi \EE_{ij} &= \partial_i \tilde{\xi}_j - \partial_j \tilde{\xi}_i 
+ \LD_\xi \EE_{ij} + \LD_{\tilde{\xi}} \EE_{ij} 
 - \EE_{ik} \left( \tilde{\partial}^k \xi^\ell - \tilde{\partial}^\ell \xi^k \right) \EE_{\ell j}.
\end{split}
\end{equation}
This is in fact quite a natural expression. The first three terms are the standard terms including the Kalb-Ramond gauge transformation and the usual Lie derivative. The last three terms are zero in a situation where the theory does not depend on the dual coordinate $\tilde x$. They are the counterparts to the first three terms which make the transformation compatible with $O(D,D)$. 
The field ${\cal E}$ appears additionally in the last terms in order to get the right index structure. Hence, all the terms that appear here are expected and natural. However, proving the gauge invariance directly is hard.

\subsection{Formulation Using the Generalized Metric}

As next step we want to arrive at an even better formulation of the action without explicit reference to the metric $g$. Ideally we want to express everything in terms of the generalized metric only, in a form that resembles the Einstein-Hilbert action as far as possible.

For example, for the dilaton we previously found the $O(D,D)$ invariant term 
$$4\DD^i d \DD_i d\,. $$ This is actually a complicated term since the ${\cal E}$ is contained in the derivatives ${\cal D}$. We can also try to formulate a dilaton term with usual partial derivatives only, but then we must be careful how to contract the indices. Certainly a contraction with $\eta$ is not reasonable, since then the constraint $\partial^M A \, \partial_M B = 0$ would kill this term. The only other possibility is to contract the indices with ${\cal H}$, yielding a term $$4{\cal H}^{MN}\partial_Md\partial_Nd\,. $$ It takes only little calculation to see that this terms is identical to the dilaton term used above. The advantage of this formulation is that we got rid of the explicit appearance of ${\cal E}$ and introduced ${\cal H}$ instead.

This does not only work for the dilaton term, but also all other terms in this action can be rephrased in this way. Doing so one finds the action
\eqn{\label{genmetaction}
	S_{\HH} = \int dx \, d \tilde{x} \, e^{-2d} \Bigl(\,
 \frac{1}{8} \,& \HH^{MN} \partial_M \HH^{KL} \partial_N \HH_{KL} 
 - \frac{1}{2} \, \HH^{MN} \partial_N \HH^{KL} \partial_L \HH_{MK} \\
 - &2 \, \partial_M d \, \partial_N \HH^{MN} 
 + 4 \, \HH^{MN} \partial_M d \, \partial_N d \Bigr) \ .
}
This action is $O(D,D)$-invariant since all indices are correctly contracted. This action is identical to the action in (\ref{S1}) although this takes some computation to verify. Finally, by dropping the $\tilde x$-dependence it reduces to the expected low-energy action (\ref{lowenergyaction}).

\subsection{Generalized Lie Derivative}

The action (\ref{genmetaction})  also comes with a gauge symmetry, and this is quite surprising and rather elegant. In a conventional setting the Lie derivatives appearing in such a theory are
\begin{align}
\LD_\xi A_M &= \xi^P \partial_P A_M + \partial_M \xi^P A_P\,,\\[0.3ex]
\LD_\xi B^N &= \xi^P \partial_P B^N - \partial_P \xi^N B^P\ .
\end{align}
In our setting here we cannot use these;  there is a very basic reason why the normal Lie derivative is not applicable. Since we include the Kalb-Ramond field in our theory, there are redundant gauge transformations where the one-form gauge parameter is $d$-exact.  
In double field theory the vector field gauge parameter can also be 
trivial.  Indeed, consider the gauge parameter  $\xi^M$ to be the derivative of some $\chi$, in components
\eqn{
	\xi^M = \begin{pmatrix} \tilde{\xi}_i \\ \xi^i \end{pmatrix} = \begin{pmatrix} \partial_i \chi \\ \tilde{\partial}^i \chi \end{pmatrix} =
	 \partial^M\chi\ .
}
The one-form $\tilde{\xi}_i$ is trivial because it is a derivative and so is the vector $\xi^i$ being a dual derivative. Hence $\xi^M$ is a trivial gauge parameter and it should generate no
Lie derivative. We see, however, that 
 \eqn{
	\LD_{\xi = \partial \chi} A_M = \partial^P \chi \partial_P A_M + \partial_M \left( \partial^P \chi \right) A_P \neq 0.
}
The first term is zero because of the constraint, but the second term is not zero. Since the Lie derivative does not vanish we should modify its definition. In fact there is a natural way to do so. Using the metric $\eta^{MN}$ it is possible to define a generalized Lie derivative~by
\eqn{
	\GLD_\xi A_M &\ \equiv \ \xi^P \partial_P A_M + \left( \partial_M \xi^P - \underline{\partial^P \xi_M} \right) A_P, \label{gld1} \\
	\GLD_\xi B^N &\ \equiv \  \xi^P \partial_P B^N - \left( \partial_P \xi^N - \underline{\partial^N \xi_P} \right) B^P\ .
}
The underlined terms are new and writing them uses the metric twice:
once to raise the derivative index and once to lower the gauge parameter index. The conventional Lie derivative distinguishes very much between covariant and contravariant indices. The generalized Lie derivative is more democratic and treats covariant and contravariant indices in a more symmetric way. It is now easy to verify that the generalized Lie derivative along a trivial field vanishes:
\eqn{
	\GLD_{\xi = \partial \chi} A_M = \partial^P \chi \partial_P A_M 
	+  \left(  \partial_M \partial^P \chi - \partial^P \partial_M \chi \right)  A_P = 0 \ .
}
$\GLD$ is the correct Lie derivative to use in our theory.  Generalized
tensors are objects with $O(D,D)$ indices $M, N, \cdots$, up or down, for which the (generalized) Lie derivative takes the form implied by (\ref{gld1}).

With the new generalized Lie derivative at hand we can now write the
gauge transformations. The gauge transformations of the
generalized metric are given by
\begin{equation}
	\delta \HH^{MN} = \GLD_\xi \HH^{MN}\ .
\end{equation}
For the dilaton we have
\begin{equation}
\delta e^{-2d} = \partial_M \left[ \xi^M e^{-2d} \right]\, .
\end{equation}
Both transformations vanish for $\xi^M = \partial^M \chi$.

The commutator of two generalized  Lie derivatives gives a very 
elegant expression
\begin{equation}
\left[ \GLD_{\xi_1}, \GLD_{\xi_2} \right] = - \GLD_{\left[ \xi_1, \xi_2 \right]_\text{C}}\ .
\label{gldcomm}
\end{equation}
The commutator is itself a generalized Lie derivative with parameter obtained by the $C$-bracket.  This shows that the $C$-bracket determines the algebra of symmetries of this theory.

\begin{Ex}
Use (\ref{gld1}) to prove that (\ref{gldcomm}) holds when acting on $A_M$.
\end{Ex}

\subsection{Generalized Einstein-Hilbert Action}

We have constructed two Lagrangians $\mathcal{L}_{\EE,d}$ and $\mathcal{L}_{\HH}$ which look very different since they are formulated in different variables, but are in fact  equal. Both are T-duality invariant, and they use field variables that reflect the doubling of coordinates. The second one, $\mathcal{L}_{\HH}$, is perhaps most novel because it completely relies on the use of the generalized metric, which is some kind of metric for a space with doubled coordinates.

Although the Lagrangian $\mathcal{L}_{\HH}$ is already written in a 
reasonably nice form, one can try to take this construction even further. One may ask if there is such a thing as a generalized Ricci curvature or a generalized scalar curvature. In fact, the answer is positive and both objects can be constructed out of the generalized metric {\it and} the dilaton. Curiously, it seems that there
is no ``generalized" Riemann curvature, although this has not been 
established for certain.  We do not need the Riemann curvature for writing down a generalized Einstein-Hilbert action, so we will leave this question aside.

The generalized scalar curvature $\RR$ is given by the expression
\eqn{\label{aoki}
\RR &= 4 \, \HH^{MN} \, \partial_M \partial_N d - \partial_M \partial_N \HH^{MN} 
  - 4 \, \HH^{MN} \, \partial_M d \, \partial_N d + 4 \, \partial_M \HH^{MN} \, \partial_N d \\
 & \qquad+ \frac{1}{8} \, \HH^{MN} \, \partial_M \HH^{KL} \, \partial_N \HH_{KL} 
  - \frac{1}{2} \, \HH^{MN} \, \partial_M \HH^{KL} \, \partial_K \HH_{NL} \ .
}
It does contain second derivatives, which is indeed expected since just like in gravity one cannot construct a scalar curvature with just one derivative.
Note that the derivatives appearing here are $\partial$ and not $\DD$, so this imposes no problem since they transform with constant $h$.
Each term in (\ref{aoki}) is $O(D,D)$ invariant, but only the full combination
of terms is a generalized scalar.

A simple rearrangement of total derivatives in $S_{\HH}$ shows that
\eqn{
	S_{\HH} = \int dx \, d \tilde{x} \, e^{-2 d} \, \RR(\HH,d) \ .
}
We see that the action takes a very simple form in terms of the
generalized scalar curvature. It looks rather analogous to the conventional Einstein-Hilbert action.

In order to prove the gauge invariance of $S_{\cal H}$ we can calculate $\delta_\xi {\cal R}$ using $\delta_\xi {\cal H}$ and $\delta_\xi d$.  
A substantial calculation confirms that  ${\cal R}$ is a generalized scalar: 
\eqn{
	\delta_{\xi} \RR = \xi^M \partial_M \RR  \ .
}
Since ${\cal R}$ is a generalized scalar and $e^{-2d}$ is a generalized density, the action is gauge invariant. 
When the dependence on $\tilde x$ is ignored (that is, setting $\tilde\partial= 0$)  the generalized scalar curvature reduces to
\eqn{
\label{eqreducedR}
	\left. \RR \right|_{\tilde{\partial} = 0} = R + 4 \left( \Box \phi - (\partial \phi)^2 \right) - \frac{1}{12} \, H^2\ ,
}
with $H=dB$ and $R$ being the conventional Ricci scalar.
This shows that scalars in general relativity do not necessarily correspond to generalized scalars in the double field theory. In general relativity all three terms on the right-hand side of (\ref{eqreducedR}) are scalars
but are not separately $O(D,D)$ invariant. In ${\cal R}$ all terms
are $O(D,D)$ invariant, but separately are not generalized scalars.

\medskip

In these lectures we have given a self-contained introduction to double field theory. We have constructed Lagrangians 
that implement T-duality more explicitly than before. We have seen the natural emergence of the Courant-bracket and how the generalized metric provides a natural variable for the formulation of the theory.  One can view the Lagrangians built here as rewritings of the familiar theory that make $O(D,D)$ symmetry manifest.  To obtain such Lagrangians we had to impose the ``strong" constraint, and it is not yet clear if  this constraint may be relaxed. This also means that the power of double field theory has not yet been fully unleashed.

\section{Acknowledgments}  I would like to thank the organizers of
the International School for their invitation to lecture and for their hospitality.
I am also grateful to Marco Baumgartl and Nicolas Moeller who prepared
an excellent version of the lecture notes that was easy to edit and finalize.
Finally, I  thank Olaf Hohm for comments and suggestions on this draft.

\end{document}